# Boosting spin-orbit torque efficiency by spin-current generator/magnet/oxide superlattices

Lijun Zhu,[1,2*] Jingwei Li,[3] Lujun Zhu,[4] Xinyue Xie[4]

1. State Key Laboratory of Superlattices and Microstructures, Institute of Semiconductors, Chinese Academy of Sciences, Beijing 100083, China
2. College of Materials Science and Opto-Electronic Technology, University of Chinese Academy of Sciences, Beijing 100049, China
3. Multi-scale Porous Materials Center, Institute of Advanced Interdisciplinary Studies & School of Chemistry and Chemical Engineering, Chongqing University, Chongqing 400044, China
4. College of Physics and Information Technology, Shaanxi Normal University, Xi'an 710062, China
*ljzhu@semi.ac.cn

**Abstract:** Efficient manipulation of magnetic materials is essential for spintronics. In conventional spin-current generator/magnet (SCG/M) bilayers, interfacial spin-orbit torques (SOTs) lose effectiveness in applications that require large magnetic layer thickness to maintain magnetic anisotropy and stability at lateral sizes of tens of nanometers (e.g. magnetic tunnel junctions and racetrack nanowires). Here, we develop an universally workable 3D spin-orbit material scheme in which the SOT efficiency can be remarkably boosted up towards infinite by stacking [SCG/M/oxide]$_n$ superlattices, with the oxide layers breaking the inversion symmetry. We demonstrate that this superlattice scheme not only promotes perpendicular magnetic anisotropy for effectively rather thick magnetic layer but also enables switching of such thick magnetic layers by interfacial SOTs with $n^2$ times lower power consumption than the corresponding conventional bilayer scheme with the same total thicknesses for the SCG and the M. In contrast, we find that spin torque diminishes in the second-type superlattices [SCG/M]$_n$ lacking inversion symmetry breaking. These results provide in-depth understanding of SOTs in magnetic multilayers and establish [SCG/M/oxide]$_n$ superlattices as advantageous bricks for development of low-power, high-stability, and high-endurance spintronic memory and computing.

## I. Introduction

Strong spin-orbit torques (SOTs)[1,2] have promise to enable fast, low-power magnetization manipulation in magnetic memory and computing technologies. In the simple case of a bilayer consisting of a spin current generator (SCG) and a magnetic layer (M) (Fig. 1(a)), a transversely polarized spin current generated by the SCG diffuses into the M and exerts interfacial SOTs on the magnetization. When the spin Hall effect (SHE) is the dominant source of the spin current (as the case of heavy metals [2-4], Bi-Sb [5,6], Bi$_x$Te$_{1-x}$ [7], CoPt [8], FeTb [9], Co-Ni-B [10] and SrIrO$_3$ [11], etc.) and when spin current relaxes within the magnetic layer via dephasing, the damping-like SOT efficiency per current density ($\xi_{DL}^j$) is given by $\xi_{DL}^j \approx T_{int}\theta_{SH}$. Here, the internal spin Hall ratio ($\theta_{SH}$) of the SCG is the product of the resistivity ($\rho_{xx}$) and the spin Hall conductivity ($\sigma_{SH}$), i.e., $\theta_{SH} = \sigma_{SH}\rho_{xx}$; $T_{int}$ is the spin transparency of the interface which determines what fraction of spin current enters the magnet.

In such conventional bilayers $\xi_{DL}^j$ is usually low particularly when $\theta_{SH}$ is small and when $T_{int}$ is far less than unity due to spin memory loss [12-18] and spin backflow [19-23] (e.g., $\xi_{DL}^j \approx 0.05$-$0.15$ for Pt/3$d$ ferromagnet [3]). A more critical limitation of the bilayer scheme is that the efficiency of the dampinglike SOT per magnetic layer thickness ($t$), $\xi_{DL}^j/t$, is inversely proportional to $t$. This degrades the effectiveness of the interfacial SOTs in spintronic applications where the M has to be thick to maintain a bulk [9,24] or shape [25,26] perpendicular magnetic anisotropy (PMA) and/or thermal stability when the lateral size scales down to tens of nanometers (e.g. magnetic tunnel junctions and racetrack nanowires).

The main idea of this work is to "accumulate" $\xi_{DL}^j$ by stacking SCG/M/oxide superlattice. Here, the SCG generates spin current that diffuses into the adjacent M, and the oxide breaks the inversion symmetry for a total torque to occur. In the SCG/M/oxide superlattice (Fig. 1(b)), the dampinglike efficiency would be enhanced by a factor of the repeat number $n$, i.e. $\xi_{DL}^j = nT_{int}\theta_{SH}$, if all repeats are identical and the thickness of the M is greater than the spin dephasing length ($l_{dp}$). When $n$ is infinite, the spin-orbit torque would, in principle, also be diverge. Any bulk SOT generated within the M layer is ignored in this model because bulk SOT is significant in the bulk limit but vanishes at small thicknesses [8,9] necessary for the superlattice scheme. However, in the superlattice [SCG/M]$_n$/SCG with no symmetry-breaking oxide layer [Fig. 1(c)], no torque is expected due to the cancellation of the spin currents from the two SCG layers sandwiching the M. In this work, we experimentally verify that $\xi_{DL}^j$ is remarkably boosted up in SCG/M/oxide superlattice using the prototype SCG Pt, 3$d$ Co, and MgO and that the SOT diminishes in Pt/Co superlattice.

## II. Sample details

For this work, we sputter-deposited Pt 2/Co 0.8/MgO 2, [Pt 2/Co 0.8/MgO 2]$_5$, [Pt 2/Co 0.8/MgO 2]$_{10}$, Pt 20/Co 0.8/MgO 2, [Pt 2/Co 0.8/Pt 2, [Pt 2/Co 0.8]$_5$, and [Pt 2/Co 0.8]$_{10}$ (here the numbers following the materials are layer thickness in nm and the subscripts represent the repeat number). Each sample was deposited on an oxidized Si substrate with a 1 nm Ta seed layer, and capped by a 2 nm MgO and a 1.5 nm Ta layer that was oxidized upon exposure to atmosphere. The Ta seed layer is very resistive and contributes negligible spin current to other layers. The layer thicknesses are estimated using the calibrated deposition rates and the deposition time. The thickness of 0.8 nm is chosen for the Co layers as a compromise between the interfacial PMA and the spin dephasing within the Co. For electrical measurements, the samples were patterned into 5×60 μm² Hall bars (Fig. 2(a)) by photolithography and ion milling. No annealing was performed on the samples.

As measured by a superconducting quantum interference device using un-patterned sample pieces that underwent the same lithography and ion milling processes as the Hall bars, the average saturation magnetization ($M_s$) of the thin Co layers is 790-1470 emu/cm³ (see Fig. 2(b) for example of [Pt 2/Co 0.8/MgO 2]$_{10}$). $M_s$ is 1275-1425 emu/cm³ for Pt 2/Co 0.8/Pt 2, [Pt 2/Co 0.8]$_5$, and [Pt 2/Co 0.8]$_{10}$. The magnetization variation of the 0.8 nm Co layers is likely due to different degree of magnetic proximity effect [27] at the Co/Pt and the Pt/Co interfaces and reduction of Curie temperature for the thin Co induced by the roughness (see below), but these $M_s$ values are within the range of literature values [27-31]. Besides the in-plane magnetized Pt 2/Co 0.8/MgO 2, all the other samples show good PMA, as indicated by the in-plane saturation field $H_k$ (Fig. 2(b)) and by out-of-plane field dependent Hall resistance hysteresis (Fig. 2(c)).

### III. Enhancement of SOT in SCG/M/oxide superlattices

To determine the SOTs, we perform harmonic Hall voltage response (HHVR) measurements [9,32-34] by carefully taking into account thermoelectric effects. To determine the current-driven damping-like SOT field ($H_{DL}$) for the PMA samples, the first and second HHVRs, $V_\omega$ and $V_{2\omega}$, are collected as a function of the small in-plane field $H_x$ (<<$H_k$) swept along the current direction, under the applied sinusoidal electric field with magnitude of $E$=33.3 kV/m. In this case, $V_\omega$ and $V_{2\omega}$ are parabolic and linear functions of $H_x$ (see Figs. 2(d) and 2(e) for the [Pt 2/Co 0.8/MgO 2]$_{10}$ and Fig. A1 in the Appendix for the [Pt 2/Co 0.8/MgO 2]$_5$), *i.e.*,

$$V_\omega \approx V_{AHE}(1-H_x^2/2H_k^2), \quad (1)$$

$$V_{2\omega} = \frac{1}{2}\frac{V_{AHE}}{H_k^2}H_{DL}H_x + \frac{V_{ANE,z}}{H_k}H_x + V_{ANE,x}. \quad (2)$$

where $V_{AHE}$ is the anomalous Hall voltage, $V_{ANE,x}$ and $V_{ANE,z}$ are the anomalous Nernst voltages due to the longitudinal and perpendicular thermal gradients. Note that $V_{2\omega}$ contains contributions that are proportional to $H_x$ from both the dampinglike torque and the perpendicular thermal gradient. Combining Eq. (1) and Eq. (2), one obtain

$$H_{DL} = -2\frac{\partial V_{2\omega}}{\partial H_x}\Big/\frac{\partial^2 V_\omega}{\partial H_x^2} - 2H_k\frac{V_{ANE,z}}{V_{AHE}}. \quad (3)$$

Here, $V_{ANE,z}$ is equal to the value of $V_{2\omega}$ when the magnetization is fully aligned along current direction by a $H_x$ that is greater than $H_k$ (Fig. 2(f)). Note that the value of $H_{DL}$ cannot be separated from a linear scaling of $-2\frac{\partial V_{2\omega}}{\partial H_x}\Big/\frac{\partial^2 V_\omega}{\partial H_x^2}$ with $E$. This suggests that the anomalous Nernst effect should be taken into account in the analysis of resistive magnetic systems with a small SOT field and a large $H_k$. The so-called "planar Hall correction" [35-39] is not applied in the out-of-plane HHVR analysis given the fact that neglecting the "correction" for the PMA samples in the out-of-plane HHVR analysis gives results that are in close accord with the in-plane HHVR results [36,40,41] and the results from measurements that do not involve the planar Hall effect (such as optical Sagnac interferometry [37], loop shift, and switching of in-plane spin-orbit torque magnetic tunnel junctions [4]). As has been experimentally known for nearly a decade [35-39], the "planar Hall correction", if significant, causes large errors for different material systems (see Appendix Note 3 for details). However, since the $V_{PHE}/V_{AHE}$ ratio for samples in this work is 0.13-0.23, the planar Hall correction, if applied, can only change the values of $H_{DL}$ and torque efficiencies by less than 20% and will not affect our conclusion. $H_{DL}$ for the in-plane sample Pt 2/Co 0.8/MgO 2 is determined from angle-dependent in-plane HHVR measurement [36] which also takes into account any thermoelectric effect (see Fig. A2 in the Appendix).

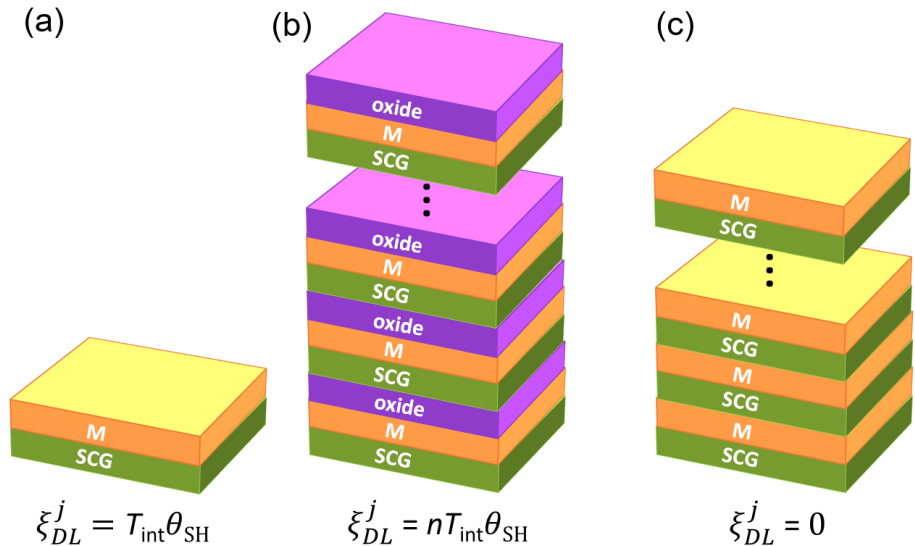

**Fig. 1.** Dampinglike spin-orbit torque efficiency expected for the three material schemes: (a) SCG/M bilayer, (b) [SCG/M/oxide]$_n$ superlattice, (c) [SCG/M]$_n$/SCG superlattice.

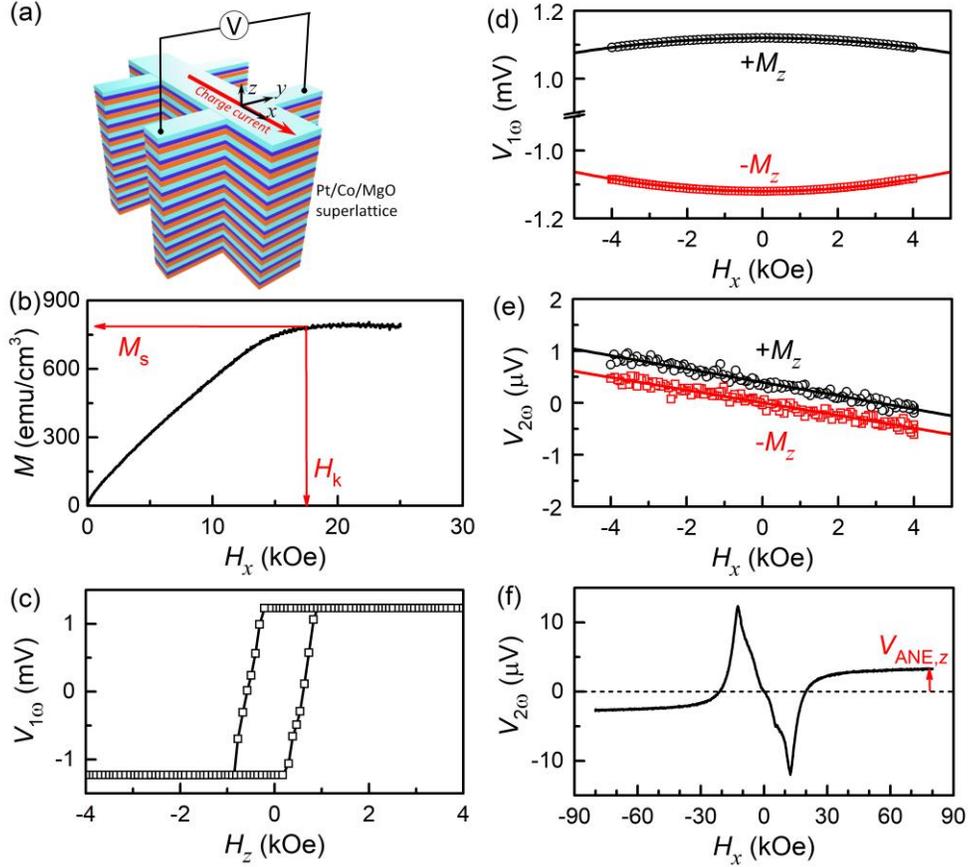

**Fig. 2**. (a) Schematic of the Pt/Co/MgO superlattice. (b) In-plane magnetization vs in-plane magnetic field ($H_x$), (c) First harmonic Hall voltage ($V_{1\omega}$) vs out-of-plane magnetic field ($H_z$), (d) $V_{1\omega}$ vs $H_x$, (e) Second harmonic Hall voltage ($V_{2\omega}$) vs $H_x$, and (e) $V_{2\omega}$ vs $H_x$ for the [Pt 2/Co 0.8/MgO 2]$_{10}$ superlattice. The red arrows in (b) indicate the saturation magnetization and perpendicular magnetic anisotropy field. Data in (d) and (e) were collected by first saturating the magnetization along film normal (i.e., $\pm M_z$) under an out-of-plane magnetic field and then sweeping $H_x$ in the range well below $H_k$ of the sample so that the magnetization is tilted in a small range and following the macrospin model. In (f), the data were collected by sweeping $H_x$ between -80 kOe to 80 kOe ($\gg H_k$), and the red arrow indicates the anomalous Nernst voltage induced by the perpendicular thermal gradient.

With the $H_{DL}$ values, the SOT efficiency per current density is calculated as

$$\xi_{DL}^j = (2e/\hbar)M_s t H_{DL}/j_c, \quad (4)$$

where $e$ is elementary charge, $\hbar$ is the reduced Plank's constant, $j_c = E/\rho_{xx}$ is the current density in the Pt layer, and $t$ is the total thickness of the magnetic layer and increases with the repeat number of the superlattice (i.e., $t$ is $n$ times the M thickness of each repeat). Strikingly, $\xi_{DL}^j$ is enhanced from 0.15 for Pt 2/Co 0.8/MgO 2, to 0.27 for five-repeat superlattice [Pt 2/Co 0.8/MgO 2]$_5$, and to 0.57 for the ten-repeat superlattice [Pt 2/Co 0.8/MgO 2]$_{10}$. $\xi_{DL}^j$ for the [Pt 2/Co 0.8/MgO 2]$_{10}$ is enhanced by a factor of $\approx 4$ compared to the Pt 2/Co 0.8/MgO and the Pt 20/Co 0.8/MgO 2 ($\xi_{DL}^j \approx 0.14$). The giant $\xi_{DL}^j$ for the [Pt 2/Co 0.8/MgO 2]$_{10}$ is also greater than the spin Hall ratio of 0.5 for Pt at resistivity of 30 μΩ cm [4]. This suggests that the SCG/M/oxide superlattice is a very effective spin-orbit material scheme to boost the total dampinglike SOT efficiency. Such enhancement cannot be attributed to the increase of the effective thickness of the Pt because we only measure a small $\xi_{DL}^j$ of 0.14 for Pt 20/Co 0.8/MgO 2 that has the same total thickness of Pt and a high magnetization of 1470 emu/cm$^3$. (we have also measured an in-plane sample Pt 20/Co 8/MgO 2 but the strong contribution of Oersted field overwhelms $V_{2\omega}$ and prevents accurate determination of $\xi_{DL}^j$). Such enhancement of $\xi_{DL}^j$ is unrelated to any spin current generation by Rashba effect or spin-orbit coupling at the interfaces because we have measured negligible SOT at interfaces of SCG/FM [38] and FM/MgO interfaces [8,16] at least in our samples.

However, the "accumulation" rate of the SOT efficiency in the [Pt/Co/MgO]$_n$ superlattice in the present work appears to be slower than expected from stacking of identical repeats. This suggests that the torque contribution of the Pt/Co/MgO repeats varies, with the bottom Pt/Co/MgO repeat grown on 1 nm Ta adhesion layer contributes the largest torque efficiency ($\approx 0.15$). First, we find that the average $\rho_{xx}$ for the Pt layers reduces from 60 μΩ cm for $n = 1$ to 35 μΩ cm for $n = 10$ (Fig. 3(b)). Here, $\rho_{xx}$ is estimated from the electrical conductance enhancement of the Pt/Co/MgO repeats relative to $n$ times the conductance of a control sample Ta1/Co 0.8/MgO 2. The resistivity reduction in the Pt layers suggests approximately a factor of 2 reduction in the average spin Hall ratio of single Pt/Co/MgO repeat because the intrinsic SHE of Pt predicts $\theta_{SH} = \sigma_{SH}\rho_{xx}$ [4]. Furthermore, the spin transparency of the Pt/Co interfaces should also decrease slightly with the repeat number due to the slowly decreasing $\rho_{xx}$. Within the Elliot-Yafet mechanism [42,43], a reduction of resistivity leads to an increase in the spin diffusion length of Pt and thus spin backflow [4,19-23] at the Pt/Co interfaces. Note that spin memory loss [12-16] that arises from interfacial spin-orbit coupling is negligible at each Pt 2/Co 0.8 interface in this work because we have measured small interfacial PMA energy density ($K_s$) that corresponds to negligible SML [16]. It has also been well established that, at interface with very weak SOC, SML is not expected whether the interface is sharp [12,13,14,16] or intermixed [44]. Within the Bruno's model [45-48], interfacial spin-orbit coupling closely scales with $K_s$ at interfaces with a relatively weak ISOC and/or a relatively weak $d$-$d$ orbital hybridization (such as Pt/Co interfaces [45]).

In addition, the variation of torque contribution of single repeat may be also related to the shortening of $l_{dp}$ of the Co for large repeat numbers. When $l_{dp}$ is greater than the layer thickness of the M, the spin current entering the M from the Pt cannot get fully relaxed before getting reflected at the oxide interface. The polarization of the reflected spin current is usually rotated relative to the incident spin current [49,50], leading to reduction of the total torque. As shown in Fig. 3(c), the average saturation magnetization for the superlattice [Pt 2/Co 0.8/MgO 2]$_n$ reduces from 1250 emu/cm$^3$ for $n$ =1 to 900 emu/cm$^3$ for $n$ = 5, and 780 emu/cm$^3$ for $n$ = 10. This would suggest a significant reduction in the average $l_{dp}$ of the Co with increasing $n$ if we assume $l_{dp}$ of the Co to scale inversely with $M_s$. The evolution of the average Co magnetization with the repeat number seems to be related to the reduction of the smoothness of the Pt/Co/MgO repeat. As indicated by the cross-sectional scanning transmission electron microscopy image of the [Pt 2/Co 0.8/MgO 2]$_{10}$ sample in Fig. 4, the Pt/Co/MgO repeat is fairly smooth for $n$ =1 but gets increasingly rough as $n$ increases towards 10. We speculate that the increased roughness could lower the Curie temperature of the thin Co layer of 0.8 nm and thus the room temperature magnetization.

While the exact cause of the resistivity and magnetization reduction needs future efforts, $\xi_{DL}^j$ of a SCG/M/oxide superlattice can be likely optimized to increase in proportion to the repeat number if the repeats can be kept identical by improved growth protocol. Alternatively, this might be also made possible if the Pt and the Co are replaced by a short-mean-free-path SCG (e.g. Au$_{0.25}$Pt$_{0.75}$ [40] or Pt$_{0.7}$(MgO)$_{0.3}$ [3]) and a magnetic layer with $M_s$ relatively insensitive to adjacent layers (e.g. FeCoB).

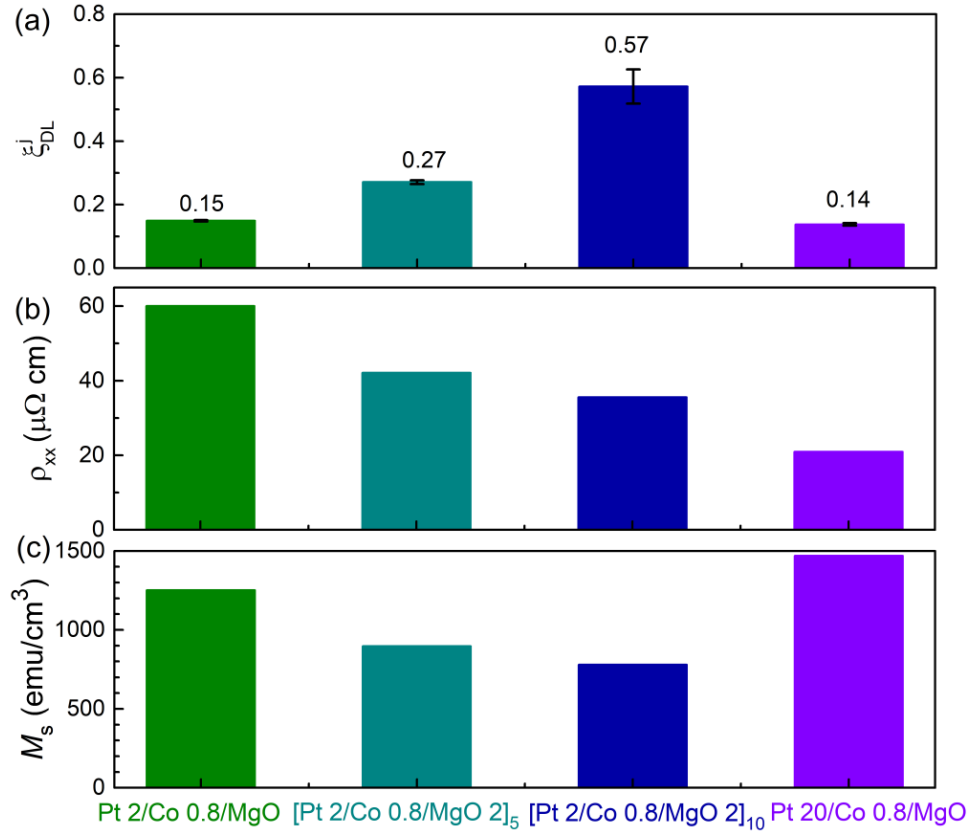

**Fig. 3**. (a) Dampinglike SOT efficiency, (b) Resistivity, and (c) Saturation magnetization for Pt 2/Co 0.8/MgO, [Pt 2/Co 0.8/MgO 2]$_5$, [Pt 2/Co 0.8/MgO 2]$_{10}$, and Pt 20/Co 0.8/MgO.

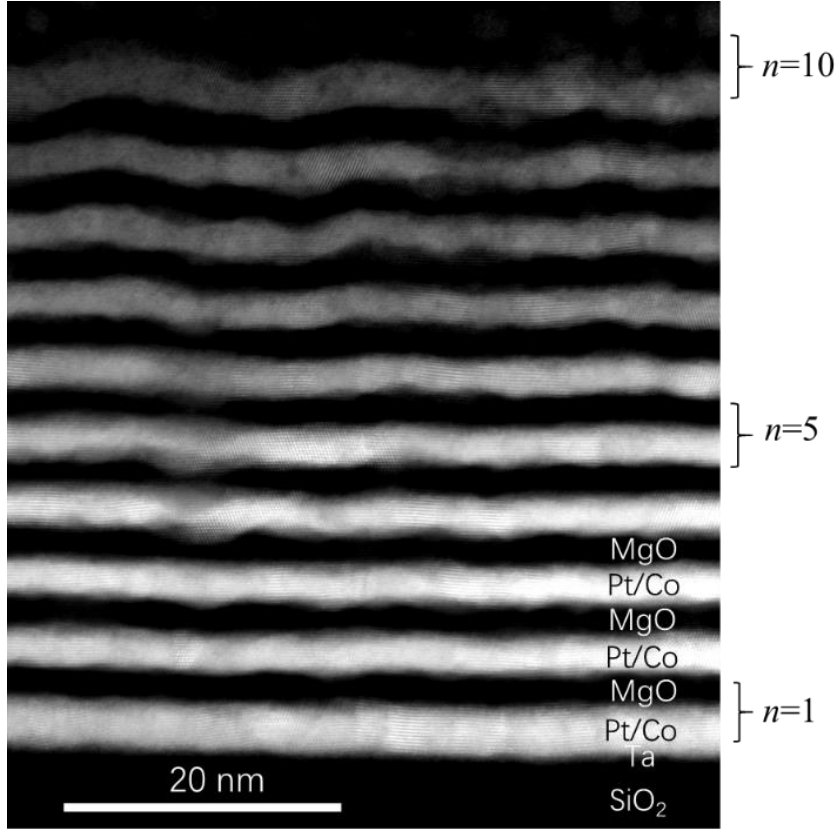

Fig. 4. Dark-field cross-sectional scanning transmission electron microscopy image for the [Pt 2/Co 0.8/MgO 2]$_{10}$ superlattice, clearly indicating reduction of smoothness with increasing repeat number $n$.

**IV. Absence of significant SOT in SCG/M superlattices**

We have also performed HHVR measurements on the Pt/Co multilayers with no MgO insertion to break the inversion symmetry. As shown in Table 1, simple stacking of Pt/Co repeats yields only small SOT efficiency because of the canceling effect of the spin currents of the two surrounding Pt layers of the Co. The non-zero SOT suggests that the Pt/Co and Co/Pt interfaces are not perfectly symmetric, with the top interface contributing greater SOT. However, this asymmetry, even for ten repeats, can only contribute a SOT efficiency that is substantially weaker than that of simple Pt/Co bilayers. Our finding of minimal SOT in Pt/Co superlattice is in sharp contrast to previous reports of giant SOTs in Co/Pd multilayers [29] and Pt/Co multilayers [28,30,31,51], but in those works the thermoelectric effect was overlooked. As we show in Table 1, if the ANE is ignored for our samples, a "giant" value of apparent $\xi_{DL}^{j}$ (such as 1.7 for [Pt 2/Co 0.8]$_{10}$) may also be concluded, which is, however, incorrect.

**Table 1**. Spin-orbit torque in Pt/Co multilayers as determined from out-of-plane HHVR with and without taking into account the anomalous Nernst effect (ANE).

| Materials | $\rho_{xx}$ ($\mu\Omega$ cm) | $\xi_{DL}^{j}$ (w/ ANE) | $\xi_{DL}^{j}$ (w/o ANE) |
|---|---|---|---|
| Pt 2/Co 0.8/Pt 2 | 57 | -0.03 | -0.08 |
| [Pt 2/Co 0.8]$_5$ | 37 | -0.04 ± 0.01 | -0.44 ± 0.01 |
| [Pt 2/Co 0.8]$_{10}$ | 34 | -0.05 ± 0.06 | -1.71 ± 0.06 |

**V. Practical impact**

Finally, we discuss the practical impact of the superlattice [SCG/M/oxide]$_n$. We first demonstrate that the new spin-orbit material scheme we develop here, 3D superlattice [SCG/M/oxide]$_n$, makes it possible to switch rather thick 3$d$ ferromagnets with strong perpendicular magnetic anisotropy. As shown in Fig. 5, the superlattice [Pt 2/Co 0.8/MgO 2]$_n$, which has a total Co thickness of 8 nm, giant perpendicular anisotropy field of $H_k$ = 17.5 kOe, high coercivity of $H_c$ = 0.6 kOe, and moderate magnetization of $M_s$ = 780 emu/cm$^3$, is deterministically switched by the SOT at a current density ($j_{c0}$) of ≈2×10$^7$ A/cm$^2$ within the Pt (the corresponding average current density is ≈1×10$^7$ A/cm$^2$ for the entire superlattice). This is in striking contrast to the conventional SCG/M bilayer scheme, for which a Co layer is unlikely to be switchable by interfacial SOT before blowout of the device due to Joule heating when the thickness is several nm, e.g. 8 nm, and can hardly maintain PMA when the effective thickness is above 1.5 nm. The ability of the superlattice to effectively switch thick magnetic layer is interesting for spintronic applications where the M has to be thick to maintain a PMA and/or thermal stability when the lateral size scales down to tens of nanometers.

More quantitatively, the current density for switching 8 nm Co in total of the superlattice is just comparable to or even smaller than that of Pt/Co and Pt alloy/Co bilayers with the Co thickness about 0.8 nm [52]. We have also recently measured $j_{c0}$ of 8.2×10$^7$ A/cm$^2$ for a Pt 2/Co 1.4 bilayer (annealed to obtain PMA)[52], predicting 4.7×10$^7$ A/cm$^2$ for a Pt 2/Co 0.8 bilayer, more than twofold higher than that for the superlattice [Pt 2/Co 0.8/MgO 2]$_n$. The current switching loop is gradual due to distribution of the depinning field (switching field) in the field switching (Fig. 2(c)) and is likely an effect of the structural smoothness variation (Fig. 4). Here we do not attempt to quantitatively estimate the value of $\xi_{DL}^j$ by applying a macrospin or a domain wall motion model to current switching experiment because of lack of simple correlation between the switching current density and $\xi_{DL}^j$ of perpendicularly magnetized SCG/M heterostructures [52].

More generally, the superlattice has $n$ times greater $\xi_{DL}^j$ and thus $n^2$ times lower power consumption to generate a given spin torque strength than the corresponding SCG/M bilayer with the same total thicknesses for the SCG and the M (assuming identical repeats for simplicity). Compared to a single SCG/M/MgO repeat, the superlattice have $n$ times smaller impedance, which makes the superlattice have a factor of $n$ reduced thermal degradation and thus improved device endurance. The superlattice also have enhanced perpendicular magnetic anisotropy (e.g., [Pt 2/Co 0.8/MgO 2]$_n$ is in-plane magnetized for $n$ =1 but perpendicularly magnetized for $n$ = 5 and 10) and thus thermal stability compared to the single SCG/M bilayer. The superlattice also provide possibility to flexibly tune the magnetic parameters (e.g. PMA and Dzyaloshinskii-Moriya interaction) via the layer thicknesses and the repeat number. Therefore, such SCG/M/oxide superlattice is a particularly advantageous spin-orbit material scheme for development of energy-efficient, high-density, high-stability, high-endurance spintronic memory and computing devices.

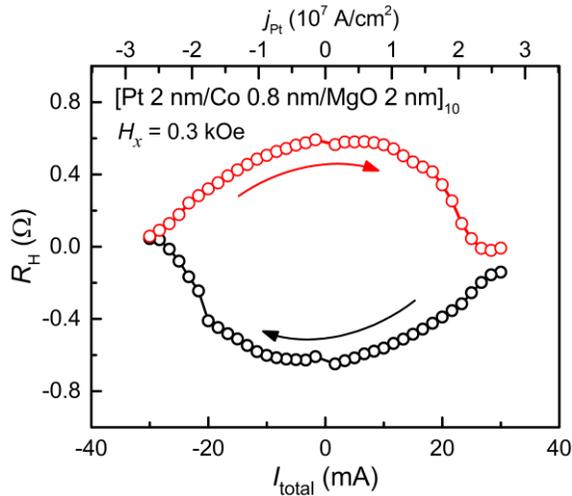

Fig. 5 Hall resistance of the superlattice [Pt 2/Co 0.8/MgO 2]$_n$ vs total current and current density in the Pt layers, showing deterministic switching of perpendicular magnetic anisotropy Co (total Co thickness = 8 nm, $H_k$=17.5 kOe, $H_c$= 0.6 kOe, $M_s$= 780 emu/cm$^3$) by the spin-orbit torque at a current density of ≈2×10$^7$ A/cm$^2$ within the Pt (the average current density of ≈1×10$^7$ A/cm$^2$ for the entire superlattice). An in-plane bias field of 0.3 kOe was applied along the current direction to overcome the DMI effect such that the spin torque can take effect.

## VI. Summary

In summary, we have developed a three-dimensional spin-orbit material scheme made by stacking spin-current generator/magnet/oxide superlattice, in which the SOT efficiency can be remarkably boosted up, even to above the spin Hall ratio of the SCG. The superlattice scheme [SCG/M/oxide]$_n$ can have enhanced perpendicular magnetic anisotropy, enhanced tunability, and $n^2$ times lower power consumption to generate a given spin torque strength than the corresponding SCG/M bilayer with the same total thicknesses for the SCG and the M. However, in the counterpart of spin-current generator/magnet superlattice with no symmetry-breaking oxide insertions the SOT diminishes. The spin-current generator/magnet/oxide material scheme, which we develop in this work, should universally work for superlattices with the SCG being a spin Hall metal (not limited to Pt), an orbital Hall metal, a topological insulator, a complex oxide, and with the magnet being a ferromagnet, a ferrimagnet, or an antiferromagnet. This novel [SCG/M/oxide]$_n$ material scheme might also be benefitted from using self-torqued magnetic layers [8,9,53-58] as the M layer if the bulk SOT can be engineered to be significant at small thicknesses (still a challenge) and additive to the interfacial SOT. These results not only advance the in-depth understanding for spin-orbit torque physics in magnetic superlattices but also establish spin-current SCG/M/oxide superlattices as new bricks for development of energy-efficient, high-endurance, high-density SOT memory and computing technologies.

**Acknowledgements**
This work was supported by the Strategic Priority Research Program of the Chinese Academy of Sciences (XDB44000000).

[1] I. M. Miron, K. Garello, G. Gaudin, P.-J. Zermatten, M. V. Costache, S. Auffret, S. Bandiera, B. Rodmacq, A. Schuhl, P. Gambardella, Perpendicular switching of a single ferromagnetic layer induced by in-plane current injection,，Nature **476**, 189 (2011).

[2] L. Liu, C. F. Pai, Y. Li, H. W. Tseng, D. C. Ralph, R. A. Buhrman, Spin-Torque Switching with the Giant Spin Hall Effect of Tantalum, Science **336**, 555 (2012).

[3] L. Zhu, L. Zhu, M. Sui, D.C. Ralph, R.A. Buhrman, Variation of the giant intrinsic spin Hall conductivity of Pt with carrier lifetime, Sci. Adv. **5**, eaav8025 (2019).

[4] L. Zhu, D. C. Ralph, R. A. Buhrman, Maximizing Spin-orbit Torque generated by the spin Hall effect of Pt, Appl. Phys. Rev. **8**, 031308 (2021).

[5] Z. Chi, Y.-C. Lau, X. Xu, T. Ohkubo, K. Hono, M. Hayashi, The spin Hall effect of Bi-Sb alloys driven by thermally excited Dirac-like electrons, Sci. Adv. **6**, eaay2324 (2020).

[6] P. N. Hai, Spin Hall Effect in Topological Insulators, J. Magn. Soc. Jpn. 44, 137 (2020).

[7] T.-Y. Chen, C.-W. Peng, T.-Y. Tsai, W.-B. Liao, C.-T. Wu, H.-W. Yen, and C.-F. Pai, Efficient Spin–Orbit Torque Switching with Nonepitaxial Chalcogenide

Heterostructures, ACS Appl. Mater. Interfaces **12**, 7788 (2020).

[8] L. Zhu, X. S. Zhang, D. A. Muller, D. C. Ralph, R. A. Buhrman, Observation of Strong Bulk Damping-Like Spin-Orbit Torque in Chemically Disordered Ferromagnetic Single Layers, Adv. Funct. Mater. **30**, 2005201 (2020).

[9] Q. Liu, L. Zhu, X.S. Zhang, D.A. Muller, D.C. Ralph, Giant bulk spin–orbit torque and efficient electrical switching in single ferrimagnetic FeTb layers with strong perpendicular magnetic anisotropy, Appl. Phys. Rev. 9, 021402 (2022).

[10] Y. Hibino, T. Taniguchi, K. Yakushiji, A. Fukushima, H. Kubota, and S. Yuasa, Large Spin-Orbit-Torque Efficiency Generated by Spin Hall Effect in Paramagnetic Co-Ni-B Alloys, Phys. Rev. Appl. **14**, 064056 (2020).

[11] T. Nan, T. J. Anderson, J. Gibbons, K. Hwang, N. Campbell, H. Zhou, Y. Q. Dong, G. Y. Kim, D. F. Shao, T. R. Paudel, N. Reynolds, X. J. Wang, N. X. Sun, E. Y. Tsymbal, S. Y. Choi, M. S. Rzchowski, Y. B. Kim, D. C. Ralph, and C. B. Eom, Anisotropic spin-orbit torque generation in epitaxial SrIrO$_3$ by symmetry design, PNAS **116**, 16186–16191 (2019).

[12] K. Chen and S. Zhang, Spin pumping in the presence of spin-orbit coupling, Phys. Rev. Lett. **114**, 126602 (2015).

[13] J. Borge and I. V. Tokatly, Ballistic spin transport in the presence of interfaces with strong spin-orbit coupling, Phys. Rev. B **96**, 115445 (2017).

[14] K. Dolui and B. K. Nikolić, Spin-memory loss due to spin-orbit coupling at ferromagnet/heavy-metal interfaces: *Ab initio* spin-density matrix approach, Phys. Rev. B **96**, 220403(R)(2017).

[15] Y. Liu, Z. Yuan, R. J. H. Wesselink, A. A. Starikov, P. J. Kelly, Interface enhancement of Gilbert damping from first principles, Phys. Rev. Lett. **113**, 207202 (2014).

[16] L. Zhu, D. C. Ralph, R. A. Buhrman, Spin-Orbit Torques in Heavy-Metal–Ferromagnet Bilayers with Varying Strengths of Interfacial Spin-Orbit Coupling, Phys. Rev. Lett. **122**, 077201 (2019).

[17] J. Bass, W. Pratt, Spin-diffusion lengths in metals and alloys, and spin-flipping at metal/metal interfaces: an experimentalist's critical review, J. Phys.: Condens. Matter **19**, 183201 (2007)

[18] J.-C. Rojas-Sánchez, N. Reyren, P. Laczkowski, W. Savero, J.-P. Attané, C. Deranlot, M. Jamet, J.-M. George, L. Vila, and H. Jaffrès, Spin Pumping and Inverse Spin Hall Effect in Platinum: The Essential Role of Spin-Memory Loss at Metallic Interfaces, Phys. Rev. Lett. **112**, 106602 (2014).

[19] L. Zhu, D. C. Ralph, R. A. Buhrman, Effective Spin-Mixing Conductance of Heavy-Metal–Ferromagnet Interfaces, Phys. Rev. Lett. **123**, 057203 (2019).

[20] L. Zhu, L. Zhu, R. A. Buhrman, Fully Spin-Transparent Magnetic Interfaces Enabled by the Insertion of a Thin Paramagnetic NiO Layer, Phys. Rev. Lett. **126**, 107204 (2021).

[21] P. M. Haney, H. W. Lee, K. J. Lee, A. Manchon, M. D. Stiles, Current induced torques and interfacial spin-orbit coupling: Semiclassical modeling, Phys. Rev. B **87**, 174411 (2013).

[22] Y.-T. Chen, S. Takahashi, H. Nakayama, M. Althammer, S. T. B. Goennenwein, E. Saitoh, and G. E. W. Bauer, Theory of spin Hall magnetoresistance, Phys. Rev. B 87, 144411 (2013).

[23] V. P. Amin and M. D. Stiles, Spin transport at interfaces with spin-orbit coupling: Phenomenology, Phys. Rev. B **94**, 104420 (2016).

[24] L. Zhu, S. Nie, K. Meng, D. Pan, J. Zhao, H. Zheng, Multifunctional $L1_0$-Mn$_{1.5}$Ga Films with Ultrahigh Coercivity, Giant Perpendicular Magnetocrystalline Anisotropy and Large Magnetic Energy Product, Adv. Mater. 24, 4547 (2012).

[25] K. Watanabe, B. Jinnai, S. Fukami, H. Sato, H. Ohno, Shape anisotropy revisited in single-digit nanometer magnetic tunnel junctions, Nat. Commun. 9, 663 (2018).

[26] B. Jinnai, K. Watanabe, S. Fukami, and H. Ohno, Scaling magnetic tunnel junction down to single-digit nanometers—Challenges and prospects, Appl. Phys. Lett. 116, 160501 (2020).

[27] L.J. Zhu, D.C. Ralph, R.A. Buhrman, Irrelevance of magnetic proximity effect to spin-orbit torques in heavy-metal/ferromagnet bilayers, Phys. Rev. B 98, 134406 (2018).

[28] P. Sethi, S. Krishnia, S.H. Li, W.S. Lew, Modulation of spin-orbit torque efficiency by thickness control of heavy metal layers in Co/Pt multilayers, J. Magn. Magn. Mater. 426, 497 (2017).

[29] M. Jamali, K. Narayanapillai, X. Qiu, L. M. Loong, A. Manchon, and H. Yang, Spin-orbit torques in Co/Pd multilayer nanowires, Phys. Rev. Lett. 111, 246602 (2013).

[30] K.-F. Huang, D.-S. Wang, H.-H. Lin, and C.-H. Lai, Engineering spin-orbit torque in Co/Pt multilayers with perpendicular magnetic anisotropy, Appl. Phys. Lett. 107, 232407 (2015).

[31] F. Xue, S.-J. Lin, M. DC, C. Bi, X. Li, W. Tsai, and S. X. Wang, Tunable spin–orbit torque efficiency in inplane and perpendicular magnetized [Pt/Co]$_n$ multilayer, Appl. Phys. Lett. 118, 042405 (2021).

[32] A. Ghosh, K. Garello, C. O. Avci, M. Gabureac, and P. Gambardella, Interface-Enhanced Spin-Orbit Torques and Current-Induced Magnetization Switching of Pd/Co/AlO$_x$ Layers, Phys. Rev. Appl. 7, 014004 (2017).

[33] H. Yang, H. Chen, M. Tang, S. Hu, X. Qiu, Characterization of spin-orbit torque and thermoelectric effects via coherent magnetization rotation, Phys. Rev. B 102, 024427 (2020).

[34] M. Hayashi, J. Kim, M. Yamanouchi, and H. Ohno, Quantitative characterization of the spin-orbit torque using harmonic Hall voltage measurements, Phys. Rev. B 89, 144425 (2014).

[35] J. Torrejon, J. Kim, J. Sinha, S. Mitani, M. Hayashi, M. Yamanouchi, H. Ohno, Interface control of the magnetic chirality in CoFeB/MgO heterostructures with heavy-metal underlayers, Nat. Commun. **5**, 4655 (2014).

[36] L. J. Zhu, K. Sobotkiewich, X. Ma, X. Li, D. C. Ralph,

R. A. Buhrman, Strong Damping-Like Spin-Orbit Torque and Tunable Dzyaloshinskii–Moriya Interaction Generated by Low-Resistivity $Pd_{1-x}Pt_x$ Alloys, Adv. Funct. Mater. **29**, 1805822 (2019).

[37] S. Karimeddiny, T. M. Cham, D. C. Ralph, Y. K. Luo, Sagnac interferometry for high-sensitivity optical measurements of spin-orbit torque, arXiv:2109.13759 (2021).

[38] L. Zhu, R.A. Buhrman, Absence of Significant Spin-Current Generation in Ti/Fe−Co−B Bilayers with Strong Interfacial Spin-Orbit Coupling, Phys. Rev. Appl. **15**, L031001 (2021).

[39] Y.C. Lau, M. Hayashi, Jpn. J. Appl. Phys. **56**, 0802B5 (2017)

[40] L. Zhu, D. C. Ralph, R. A. Buhrman, Highly Efficient Spin-Current Generation by the Spin Hall Effect in $Au_{1-x}Pt_x$, Phys. Rev. Appl. **10**, 031001(2018).

[41] L. Zhu, L. Zhu, S. Shi, M. Sui, D.C. Ralph, and R.A. Buhrman, Enhancing Spin-Orbit Torque by Strong Interfacial Scattering From Ultrathin Insertion Layers, Phys. Rev. Appl. **11**, 061004 (2019).

[42] R.J. Elliott, Theory of the Effect of Spin-Orbit Coupling on Magnetic Resonance in Some Semiconductors, Phys. Rev. **96**, 266 (1954).

[43] Y. Yafet, g Factors and Spin-Lattice Relaxation of Conduction Electrons, Solid State Phys. **14**, 1 (1963).

[44] L. Zhu, D.C. Ralph, R.A. Buhrman, Enhancement of spin transparency by interfacial alloying, Phys. Rev. B **99**, 180404 (2019)

[45] L. Zhu, L. Zhu, X. Ma, X. Li, R. A. Buhrman, Critical role of orbital hybridization in the Dzyaloshinskii-Moriya interaction of magnetic interfaces, Comm. Phys. **5**, 151 (2022).

[46] D. Weller, J. Stöhr, R. Nakajima, A. Carl, M. G. Samant, C. Chappert, R. Mégy, P. Beauvillain, P. Veillet, and G. A. Held, Microscopic Origin of Magnetic Anisotropy in Au/Co/Au Probed with X-Ray Magnetic Circular Dichroism, Phys. Rev. Lett. **75**, 3752 (1995).

[47] P. Bruno, Tight-binding approach to the orbital magnetic moment and magnetocrystalline anisotropy of transition-metal monolayers, Phys. Rev. B **39**, 865 (1989).

[48] C. Andersson, et al. Influence of ligand states on the relationship between orbital moment and magnetocrystalline anisotropy, Phys. Rev. Lett. **99**, 177207 (2007).

[49] Z. Luo, Q. Zhang, Y. Xu, Y. Yang, X. Zhang, and Y. Wu, Spin-Orbit Torque in a Single Ferromagnetic Layer Induced by Surface Spin Rotation, Phys. Rev. Applied **11**, 064021(2019).

[50] Y. Ou, C.-F. Pai, S. Shi, D. C. Ralph, and R. A. Buhrman, Origin of fieldlike spin-orbit torques in heavy metal/ferromagnet/oxide thin film heterostructures, Phys. Rev. B **94**, 140414(R)(2016).

[51] B. Jinnai, C. Zhang, A. Kurenkov, M. Bersweiler, H. Sato, S. Fukami, and H. Ohno, Spin-orbit torque induced magnetization switching in Co/Pt multilayers, Appl. Phys. Lett. **111**, 102402 (2017).

[52] L. Zhu, D.C. Ralph, R.A. Buhrman, Lack of Simple Correlation between Switching Current Density and Spin-Orbit-Torque Efficiency of Perpendicularly Magnetized Spin-Current-Generator–Ferromagnet Heterostructures, Phys. Rev. Appl. **15**, 024059 (2021).

[53] L. Zhu, D.C. Ralph, R. A. Buhrman, Unveiling the Mechanism of Bulk Spin-Orbit Torques within Chemically Disordered $Fe_xPt_{1-x}$ Single Layers, Adv. Funct. Mater. **31**, 2103898 (2021).

[54] Y. Cao, Y. Sheng, K. W. Edmonds, Y. Ji, H. Zheng, K. Wang, Deterministic Magnetization Switching Using Lateral Spin–Orbit Torque, Adv. Mater. **32**, 1907929 (2020).

[55] M. Tang, K. Shen, S. Xu, H. Yang, S. Hu, W. Lü, C. Li, M. Li, Z. Yuan, S. J. Pennycook, K. Xia, A. Manchon, S. Zhou, X. Qiu, Bulk Spin Torque-Driven Perpendicular Magnetization Switching in $L1_0$ FePt Single Layer, Adv. Mater. **32**, 2002607 (2020).

[56] J. W. Lee, J. Y. Park, J. M. Yuk, B.-G. Park, Spin-Orbit Torque in a Perpendicularly Magnetized Ferrimagnetic Tb-Co Single Layer, Phys. Rev. Appl. **13**, 044030 (2020).

[57] L. Liu, J. Yu, R. González-Hernández, C. Li, J. Deng, W. Lin, C. Zhou, T. Zhou, J. Zhou, H. Wang, R. Guo, H. Y. Yoong, G. M. Chow, X. Han, B. Dupé, J. Železný, J. Sinova, J. Chen, Electrical switching of perpendicular magnetization in a single ferromagnetic layer, Phys. Rev. B **101**, 220402 (2020).

[58] Z. Zheng, Y. Zhang, L. V. Lopez-Dominguez, Sánchez-Tejerina, J. Shi, X. Feng, L. Chen, Z. Wang, Z. Zhang, K. Zhang, B. Hong, Y. Xu, Y. Zhang, M. Carpentieri, A. Fert, G. Finocchio, W. Zhao, P. K. Amiri, Field-free spin-orbit torque-induced switching of perpendicular magnetization in a ferrimagnetic layer with a vertical composition gradient, Nat. Comm.**12**, 4555 (2021).

**Appendix Note 1. The [Pt 2/Co 0.8/MgO 2]$_5$ superlattices**

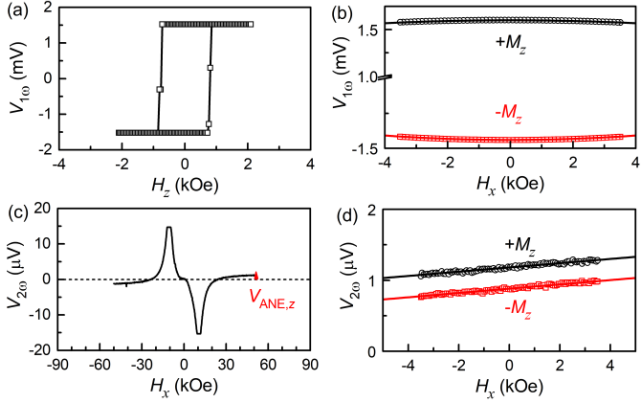

Fig. A1. Out-of-plane harmonic Hall voltage response measurement on the [Pt 2/Co 0.8/MgO 2]$_5$ superlattices. (a) First harmonic Hall voltage ($V_{1\omega}$) vs out-of-plane magnetic field ($H_z$), (b) $V_{1\omega}$ vs $H_x$, (c),(d) Second harmonic Hall voltage ($V_{2\omega}$) vs $H_x$. The red arrow in (c) indicates the anomalous Nernst voltage induced by the perpendicular thermal gradient. The solid lines in (b) and (d) represent the fits of the data to Eq. (1) and Eq. (2) in the main text, respectively.

**Appendix Note 2. In-plane magnetized Pt 2/Co 0.8/MgO 2**

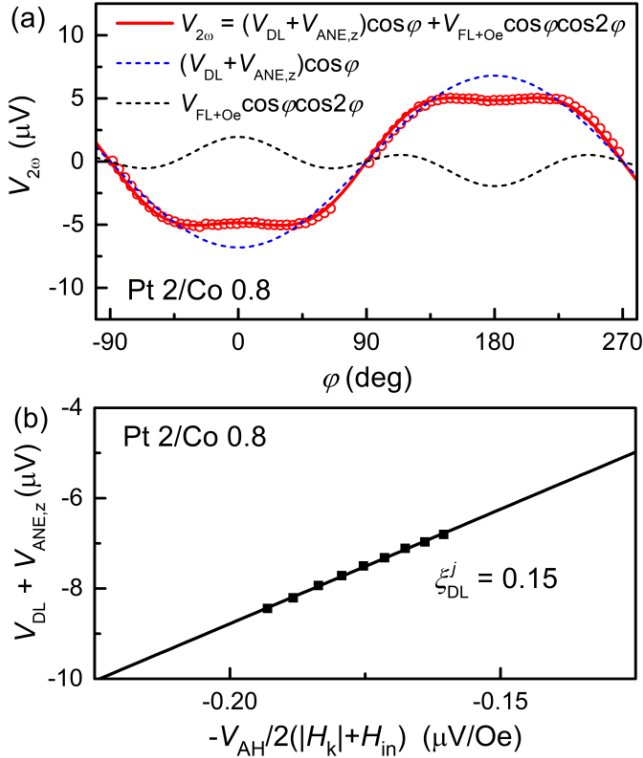

Fig. A2. (a) The second harmonic Hall voltage response of the in-plane magnetized Pt 2/Co 0.8/MgO 2. The red curve refers to the best fits of the data to Eq. (A1). (b) Linear fit of $V_{DL}$ vs $-V_{AH}/2(H_{in}+|H_k|)$, the slope pf which yields the value of $H_{DL}$. The applied electric field is 66.7 kV/m.

The dampinglike torque effective field $H_{DL}$ of the in-plane magnetized Pt 2/Co 0.8/MgO 2 are determined from in-plane harmonic Hall voltage response (HHVR) measurements. As shown in Fig. A2(a), the second HHVR of an in-plane magnetization is given by

$$V_{2\omega} = (V_{DL} + V_{ANE,Z})\cos\varphi + V_{FL+Oe}\cos\varphi\cos 2\varphi, \quad (A1)$$

where $V_{DL} = -V_{AH}H_{DL}/2(H_{in}+|H_k|)$ is the second HHVR from the dampinglike SOT, $V_{FL}$ the second HHVR from the fieldlike SOT and Oersted field torque, $V_{ANE}$ the anomalous Nernst voltage, and $H_{in}$ the in-plane bias field. The slope of $V_{DL}$ vs $-V_{AH}/2(H_{in}+|H_k|)$ yields the value of $H_{DL}$ (Fig. A2(b)).

**Appendix Note 3.** Absence of the "planar Hall correction"

As initially suggested by Hayashi *et al*. in 2014 [34], the planar Hall effect appears to be involved in the out-of-plane HHVR on samples with perpendicular magnetic anisotropy (PMA) and to modify the fieldlike and dampinglike spin-torque efficiencies as

$$\xi_{DL(FL)}^j(\text{with correction}) = (\xi_{DL(FL)}^j + 2\zeta\xi_{FL(DL)}^j)/(1-4\zeta^2), \quad (A2)$$

where $\zeta = V_{PHE}/V_{AHE}$, and $V_{PHE}$ and $V_{AHE}$ are the planar Hall voltage and the anomalous Hall voltage, respectively.

However, since the same year 2014 this so-called "planar Hall correction" has been known to cause errors by researchers including the developer of this "correction" [35-39]. First, this "correction" leads to infinite estimates for spin-orbit torque efficiencies when $V_{PHE}/V_{AHE} = 0.5$ (see Eq. (A2)). Experimentally, when $V_{PH}/V_{AH}$ is very large, this "correction" leads to unrealistic magnitudes and even sign reversals for the extracted values of $\xi_{DL}^j$, with large discrepancies compared to other measurement methods (see Table A1 for examples on different material systems).

In contrast, *neglecting the "correction" for the PMA samples* in the out-of-plane HHVR measurement consistently gives results that are in close accord with the in-plane HHVR results from in-plane magnetic anisotropy samples with the same heavy metal/ferromagnet components and the similar resistivities and thicknesses (see Table A1, Fig. A3, and Refs. [36,40,41]) and with results from measurements that do not involve the planar Hall effect (such as optical Sagnac interferometry [37], domain wall motion [35], loop shift, and switching of in-plane spin-orbit torque magnetic tunnel junctions [4]). More discussions have been reported in the supplementary Materials in [16,35,36]. Therefore, *it has been experimentally clear that the so-called "planar Hall correction" should be avoided.*

Therefore, the "planar Hall correction" is not applied to the Pt/Co/MgO or Pt/Co superlattices in the main text (Fig. 3 and Table 1). The physical mechanism for the absence of the planar Hall correction in the out-of-plane HHVR measurements is beyond the scope of our present work, because it is a long-standing open question in the whole spintronic community and because it is not specific to the superlattice samples in this work. We do note that, since the $V_{PHE}/V_{AHE}$ ratio for samples in this work is 0.13-0.23, the planar Hall correction, if applied, can only change the torque efficiencies by less than 20% and will not affect the conclusion of this work.

**Table A1**. The out-of-plane HHVR results of the PMA heavy metal/ferromagnet bilayer samples without and with the "planar Hall correction" vs the in-plane HHVR results on in-plane magnetized samples with the similar heavy-metal resistivities and the same ferromagnetic layer. The PMA results for $\xi_{DL}^j$ are in good agreement with in-plane HHVR results only if the "correction" is not applied. Applying the "planar Hall correction" gives unrealistic numbers for the fieldlike and/or dampinglike torque efficiencies and alters the sign of the dampinglike torque of the Pd 4/Co 0.64 and the sign of both dampinglike and fieldlike torque of the W 2.5/CoFeB 1. These results reveal that the "planar Hall correction" should be avoided in the out-of-plane HHVR results.

| PMA samples | $V_{PH}/V_{AH}$ | PMA sample No "correction" | | PMA sample with "correction" | | In-plane sample | | Reference |
|---|---|---|---|---|---|---|---|---|
| | | $\xi_{DL}^j$ | $\xi_{FL}^j$ | $\xi_{DL}^j$ | $\xi_{FL}^j$ | $\xi_{DL}^j$ | $\xi_{FL}^j$ | |
| W 2.2/CoFeB 1 | 0.486 | -0.132 | -0.064 | -3.52 | -3.25 | - | - | [35] |
| W 2.5/CoFeB 1 | 0.54 | -0.15 | -0.005 | 0.93 | 1.00 | - | - | [35] |
| Pt 4/Co 0.75 | 0.31 | 0.21 | -0.049 | 0.29 | 0.13 | 0.19 | -0.046 | [39] |
| Pd 4/Co 0.64 | 0.56 | 0.07 | -0.050 | -0.1 | -0.16 | 0.06 | -0.0002 | [36] |
| Au$_{0.25}$Pt$_{0.75}$ 4/Co 0.8 | 0.33 | 0.30 | -0.12 | 0.39 | -0.14 | 0.32 | -0.020 | [40] |

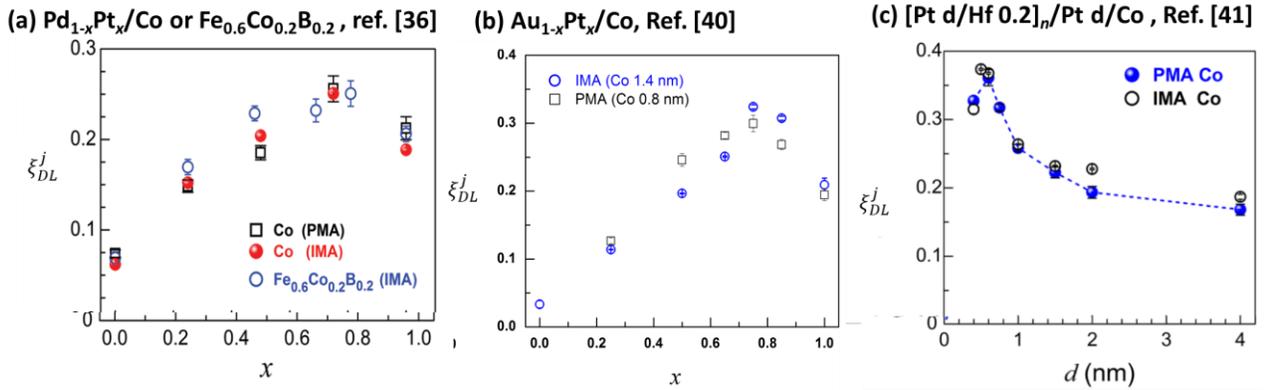

Fig. A3 Consistence of in-plane HHVR results and out-of-plane HHVR results of dampinglike spin-orbit torque efficiencies per unit current density for (a) Pd$_{1-x}$Pt$_x$/(Co or Fe$_{0.6}$Co$_{0.2}$B$_{0.2}$) bilayer with different Pt concentration in the Pd$_{1-x}$Pt$_x$ [36], (b) Au$_{1-x}$Pt$_x$ /Co bilayer with different Pt concentration in the Au$_{1-x}$Pt$_x$ [40], and (c) [Pt $d$/Hf 0.2]$_n$/Pt $d$/Co with different thickness of each Pt slide [41]. The "IMA" represents the in-plane harmonic Hall voltage response results from in-plane magnetized samples, while the "PMA" represents the out-of-plane harmonic Hall voltage response results from perpendicular magnetic anisotropy samples, with NO so-called "planar Hall correction". Clearly, the out-of-plane HHVR results without "planar Hall correction" are reasonably close to that of in-plane HHVR results. Note that, for most PMA samples in (a)-(c), the "planar Hall correction" leads to unrealistic values for the dampinglike torque efficiencies.